\documentclass[a4paper,11pt]{article}
\usepackage{amsmath}
\usepackage{amsfonts}
\usepackage{color} 
\usepackage{epsfig}
\usepackage{mathtools}
\usepackage{amsthm}

\usepackage{mathabx}

\begin{document}

\newtheorem{theorem}{Theorem}
\newtheorem{proposition}{Proposition}
\newtheorem{example}{Example}
\newtheorem{remark}{Remark}

\centerline{\begin{Large} On the conformal transformation \end{Large}}
\centerline{\begin{Large} between two anisotropic fluid spacetimes \end{Large}}

\bigskip

\centerline{Jaros\l aw Kopi\'nski\footnote{jkopinski@cft.edu.pl}}

\centerline{Center for Theoretical Physics, Polish Academy of Sciences}
\centerline{Aleja Lotnik\'ow 32/46, 02-668 Warsaw, Poland}
\abstract{ \noindent We investigate the necessary conditions for the two spacetimes, which are solutions to the Einstein field equations with an anisotropic matter source, to be related to each other by means of a conformal transformation. As a result, we obtain that if one of such spacetimes is a generalization of the Robertson-Walker solution with vanishing acceleration and vorticity, then the other one has to be in this class as well, i.e. the conformal factor will be a constant function on the hypersurfaces orthogonal to the fluid flow lines. The evolution equation for this function appears naturally as a direct consequence of the conformal transformation of the Ricci tensor.}
\section*{Introduction}
The question of whether there exist a conformal class of metrics in which there are two non-isometric Einstein metrics dates back to Brink- \\mann \cite{brinkmann}. He posed and answered it in dimension four in every signature. A similar problem arises in the conformal cyclic cosmology. According to this paradigm, the history of the universe consists of a sequence of aeons, each one being a solution of the Einstein field equations with the positive cosmological constant \cite{penrose}. The matching of two consecutive aeons is performed on a spacelike hypersurface with the vanishing Weyl tensor, which corresponds to the future null infinity of the previous aeon and the past null infinity of the next aeon. According to the reciprocity hypothesis put forward by Penrose, the metric tensors of these spacetimes are related to each other by means of conformal transformation,
\begin{equation}
\widecheck{g}_{ab} = \widehat{\Omega}^{-4} \widehat{g}_{ab}.
\end{equation}
The conformal factor $\widehat{\Omega}$ is not defined by the data from each aeon. It satisfies certain postulated equation instead, like the Yamabe-type one originally proposed by Penrose. In the recent approaches to understanding the matching of perfect fluid spacetimes with the Robertson–Walker metric and its generalizations \cite{nurowski, nurowski_meissner, tod, newman, nurowski2}, this function has been chosen to depend only on the time measured by the observers comoving with the fluid.

This work explores the question of conditions that have to be imposed on the conformal factor if two metrics $\widehat{g}_{ab}$ and $\widecheck{g}_{ab}$ are solutions of the Einstein field equations with cosmological constants and anisotropic matter sources and are related to each other by means of a conformal transformation. We consider here the simplest generalizations of the Robertson-Walker spacetime for which the Ricci tensor of one of the metrics lies in the (possibly degenerate) subclass $[S_1 - S_2 - S_3 -T ]_{(1111)} $ (alternatively $[111,1]$) of the Churchill-Pleba\'nski classification of tensors. 

As a result, we prove that if one of the solutions of the Einstein field equations has a fluid source with vanishing acceleration and vorticity and with constant density and isotropic pressure on the hypersurfaces orthogonal to its four-velocity, then the other one has those properties as well. In that case, the conformal factor varies only along the fluid flow lines. Moreover, its evolution equation can be written down as a direct consequence of the transformation relation between the timelike-timelike components of the Ricci tensors of $\widehat{g}_{ab}$ and $\widecheck{g}_{ab}$.

The structure of this article is as follows. We start with the assumption that the metric $\widehat{g}_{ab}$ is a particular solution of the Einstein field equations with the anisotropic fluid source. Then we move to the derivation of the necessary conditions on the Cotton and Bach tensors (conformal Cotton and Bach equations) that have to be satisfied in order for the metric $\widecheck{g}_{ab}$ to be in the conformal class of  $\widehat{g}_{ab}$. In the last section, we make an additional assumption that $\widecheck{g}_{ab}$ also satisfies the Einstein field equations with the source of similar type and that the fluids four velocities are related to each other in the natural way stemming from the relation between the metric tensors. As a consequence of this assumption, we obtain a condition that the conformal factor has to be a constant function on the hypersurfaces orthogonal to the fluid flow lines and derive an evolution equation for this function.

\subsection*{Notation and conventions}
We will follow the same conventions concerning various tensors definitions as in \cite{nurowski_leistner}. In particular, we will use the abstract index notation. The symmetrization and antisymmetrization brackets will be denoted by $(\cdot )$ and $[\cdot]$ respectively, i.e.
\begin{equation}
T_{(ab)} = \frac{1}{2} \left(T_{ab} + T_{ba} \right), \quad T_{[ab]} = \frac{1}{2} \left(T_{ab} - T_{ba} \right).
\end{equation} \\
Let $R_{abcd}$ be the Riemann tensor associated to the metric $g_{ab}$. It satisfies
\begin{equation} \label{riemdef}
2 \nabla_{[a} \nabla_{b]} v_c = R_{abc}{}^d v_d 
\end{equation}
for any one-form $v_c$. The Ricci tensor is defined as a partial trace of the Riemann tensor, $R_{ab}=R^c{}_{acb}$, and the scalar curvature is $R=R^{a}{}_a = R^{ab}{}_{ab}$. The traceless part of $R_{abcd}$ is called the Weyl tensor, which will be denoted as $C_{abcd}$. The following decomposition holds
\begin{equation}
R_{abcd} = C_{abcd} + 2 \left(g_{c[a}P_{b]d} + g_{d[b}P_{a]c}\right) ,
\end{equation}
where $P_{ab}$ is the Schouten tensor,
\begin{equation}
P_{ab} = \frac{1}{2}R_{ab} - \frac{R}{12} g_{ab}.
\end{equation}
It can be used to define the Cotton ($A_{abc}$) and Bach ($B_{ab}$) tensors,
\begin{equation}
\begin{aligned}
A_{abc} & = 2 \nabla_{[b}P_{c]a}, \\
B_{ab} & = - \nabla^c A_{abc} + P^{dc} C_{dacb}.
\end{aligned}
\end{equation} 
The differential Bianchi identity $\nabla_{[a}R_{bc]de} = 0$ provides a relation between the Cotton and Weyl tensors,
\begin{equation}
A_{abc}= \nabla^d C_{dabc},
\end{equation}
which highlights the trace-free nature of the former.\\
Any timelike unit vector defines a decomposition of the Weyl tensor into its electric and magnetic parts, $E_{ab}$ and $H_{ab}$ respectively,
\begin{align} \label{ebweyl}
E_{ab} = u^c u^d C_{acbd}, \quad H_{ab} = \frac{1}{2} u^c u^d \eta_{ackl}C^{kl}{}_{bd}.
\end{align}
where $\eta_{abcd}$ is the covariant Levi-Civita tensor. They uniquely define the Weyl tensor, i.e. $E_{ab} = H_{ab} =0 $ iff $C_{abcd} = 0$. Moreover,
\begin{equation} \label{ebweyl2}
C_{abcd} u^d = 2 u_{[a} E_{b]c} - \eta_{ab d } H_c{}^d,
\end{equation}
where $\eta_{abc} = u^d \eta_{dabc}$.
\section*{Anisotropic fluid spacetime}
The Einstein field equations relate the geometry of a four-dimensional Lorentzian manifold $M$ to its matter content. They read
\begin{equation}
R_{ab} - \frac{1}{2} R g_{ab} + \Lambda g_{ab} = T_{ab},
\end{equation}
where $R_{ab}$ is the Ricci tensor of the metric $g_{ab}$, $\Lambda$ is the cosmological constant and $T_{ab}$ is the energy-momentum tensor. In the sequel we will assume that the matter content is of the anisotropic fluid type with an anisotropic pressure $\pi_{ab}$, i.e. 
\begin{equation}
T_{ab} = \left( \rho + p \right) u_a u_b + p g_{ab} + \pi_{ab},
\end{equation}
where $u^a \pi_{ab} =g^{ab}\pi_{ab} =0 $. The quantities $\rho$ and $p$ can be interpreted as a density and (isotropic) pressure of the fluid, whereas $u^a$ is a timelike unit vector field (four-velocity of the fluid). \\
There exists a standard decomposition of the covariant derivative of $u^a$ into the vorticity $\omega_{ab}$, shear $\sigma_{ab}$, expansion $\theta$ and acceleration $\dot{u}_a=\nabla_u u_a$,
\begin{equation}
\nabla_a u_b = \omega_{ab} + \sigma_{ab} + \frac{1}{3} \theta \left(g_{ab} + u_a u_b\right) - u_a \dot{u}_b,
\end{equation}
where $u^b \omega_{ab} = u^b \sigma_{ab} = g^{ab} \sigma_{ab} = u^a \dot{u}_a=0$ and $\omega_{(ab)} = \sigma_{[ab]}=0$. We will denote the distribution orthogonal to $u^a$ as $\Sigma$. The orthogonal projector onto $\Sigma$ can be defined as 
\begin{equation}
h_{a}{}^{b} \vcentcolon = \delta_{a}{}^{b} + u_a u^b. 
\end{equation}

The energy-momentum tensor has to satisfy the continuity equation,
\begin{equation}
\nabla_a T^{ab} =0,
\end{equation}
which in the current scenario can be decomposed into the components along the fluid flow lines and orthogonal to them. They read
\begin{align}
\nabla_u \rho  + \left( \rho + p \right) \theta + \sigma_{ab} \pi^{ab} =0, \label{conteq1} \\
\left( \rho + p \right) \dot{u}_a + h_a{}^b \left(\nabla_b p+ \nabla_c \pi^c{}_b \right) =0. \label{conteq2}
\end{align}
Equation (\ref{conteq1}) can be treated as an expression of the second law of thermodynamics \cite{ellis}. As a result, the phenomenological equation
\begin{equation}
\pi_{ab} = - \lambda (\rho,p) \sigma_{ab},
\end{equation}
will be imposed to keep the rate of generation of entropy positive, where $\lambda>0$ is called the viscosity parameter. It is worth to notice that if $\lambda = 0$, then the energy-momentum tensor and, through the Einstein field equations, the Ricci tensor lies in the special degenerate algebraic class $[3S-T]_{(11)}$ of the Plebański-Churchill algebraic classification of tensors and has one timelike and one spacelike eigenvalue \cite{plebanski, churchill}.

The identity (\ref{riemdef}) can be applied to the four-velocity vector $u^a$ to obtain the propagation equations for $\theta$ and $\omega_{ab}$,
\begin{align}
\nabla_u \theta + \frac{1}{3}\theta^2 - \nabla_a \dot{u}^a + \sigma_{ab}\sigma^{ab} -\omega_{ab} \omega^{ab} + \frac{1}{2} \left( \rho + 3 p \right) - \Lambda =0, \label{propth} \\
h_a{}^b \nabla_u \left( \eta_b{}^{cd} \omega_{cd} \right) + \eta_a{}^{bc} \left( \frac{2}{3} \theta \omega_{bc}+ \nabla_{b} \dot{u}_c \right)-\sigma_{ab} \eta^{bcd}\omega_{cd}=0,
\end{align}
the constraint equation
\begin{equation} \label{spatth}
h_a{}^b \left(\nabla_c \omega_b{}^c - \nabla_c \sigma_b{}^c + \frac{2}{3} \nabla_b \theta   \right) + \left( \omega_{ab} + \sigma_{ab}\right) \dot{u}^a =0
\end{equation}
and the definitions of the electric and magnetic parts of the Weyl tensor in terms of the quantities describing the flow generated by $u^a$.
\subsection*{Acceleration-free flow with orthogonal hypersurfaces of constant matter density and isotropic pressure}
In the sequel we will assume that $(\widehat{M}, \widehat{g}_{ab})$ is a solution of the Einstein field equations with an anisotropic fluid source described by the ``hatted" quantities. In order to single out $\widehat{\pi}_{ab}$ ($\widehat{\sigma}_{ab}$) as the sole source of the anisotropy we will make an additional assumption concerning the matter density $\widehat{\rho}$ and isotropic pressure $\widehat{p}$,
\begin{equation}
\widehat{D}_a \widehat{\rho} = 0 = \widehat{D}_a \widehat{p} , 
\end{equation}
where $\widehat{D}_{a}$ is the Levi-Civita connection on $\widehat{\Sigma}$. As a consequence, $\widehat{D}_a \widehat{\lambda} = 0$. We will also assume that the acceleration and vorticity of $\widehat{u}^a$ vanish,
\begin{equation}
\dot{\widehat{u}}_a =0  =\widehat{\omega}_{ab} .
\end{equation} 
The condition $\widehat{\omega}_{ab} =0$ is equivalent to the assumption that $\widehat{\Sigma}$ is an integrable distribution, i.e. there exists a foliation of the spacetime $\widehat{M}$ with three-dimensional spatial leaves  (orthogonal to $\widehat{u}^a$). Those assumptions were chosen in a way which makes $(\widehat{M}, \widehat{g}_{ab})$ a simplest generalization of the Robertson-Walker spacetime, i.e. its Ricci tensor lies in the $[S_1 - S_2 - S_3 -T]_{(1111)}$ class of the Pleba\'nski-Churchill classification.

In this current setting, the spatial part of the matter continuity equation (\ref{conteq2}) implies
\begin{equation} \label{divsig}
\widehat{D}_a \widehat{\sigma}^a{}_b =0.
\end{equation}
Moreover, constraint (\ref{spatth}) gives
\begin{equation}
\widehat{D}_a \widehat{\theta} =0,
\end{equation}
where (\ref{divsig}) has been used, i.e. the expansion $\widehat{\theta}$ is a constant function on $\widehat{\Sigma}$. This fact can be combined with the spatial derivative of the timelike projection of the matter continuity equation, (\ref{conteq1}), to get
\begin{equation}
\widehat{D}_a \widehat{\sigma}_{bc} \widehat{\sigma}^{bc} =0.
\end{equation}
Those considerations can be summarized as
\begin{proposition} \label{prop1} 
Let $(\widehat{M}, \widehat{g}_{ab})$ be an anisotropic fluid spacetime with the fluid four-velocity having vanishing acceleration and vorticity. Let the matter density $\widehat{\rho}$ and isotropic pressure $\widehat{p}$ be constant functions on the hypersurface orthogonal to the fluid flow lines $\Sigma$. Then
\begin{equation} 
\widehat{D}_b \widehat{\sigma}^b{}_a =\widehat{D}_a \widehat{\theta} =  \widehat{D}_a \widehat{\sigma}_{bc} \widehat{\sigma}^{bc} = 0,
\end{equation}
where $\widehat{D}_a$ is the Levi-Civita connection on $\widehat{\Sigma}$, i.e. the shear tensor is divergence-free on $\widehat{\Sigma}$, whereas its norm and the expansion scalar are constant functions on this hypersurface.
\end{proposition}
We will move now to the properties of the Cotton and Bach tensors of $(\widehat{M}, \widehat{g}_{ab})$. A direct computation yields the following result.
\begin{theorem} \label{th1}
Let $(\widehat{M}, \widehat{g}_{ab} )$ be a four-dimensional spacetime with a timelike unit vector field $\widehat{u}^a$ with vanishing acceleration $\dot{\widehat{u}}_a$ and vorticity $\widehat{\omega}_{ab}$. Let the matter density $\widehat{\rho}$ and isotropic pressure $\widehat{p}$ be constant functions on the hypersurfaces $\widehat{\Sigma}$ orthogonal to $\widehat{u}^a$, i.e. $\widehat{D}_a \widehat{\rho} = \widehat{D}_a \widehat{p} = 0$. Then the following projections of the Cotton and Bach tensors vanish,
\begin{align}
\widehat{u}^a \widehat{A}_{abc} = 0, \\
\widehat{h}_c{}^b \widehat{u} ^a \widehat{B}_{ab} = 0.
\end{align} 
\end{theorem}
\noindent
A similar theorem has been obtained by Leistner and Nurowski for the pure radiation spacetimes with parallel rays spanned by null vectors \cite{nurowski_leistner}. 
\begin{remark}
We have
\begin{equation}
\begin{split}
\widehat{E}_{ab} &  = -\frac{\widehat{\lambda}}{2} \widehat{\sigma}_{ab} -\widehat{\sigma}_{ac} \widehat{\sigma}_{b}{}^c -\frac{2}{3} \widehat{\theta} \widehat{\sigma}_{ab} -\widehat{\nabla}_{\widehat{u}} \widehat{\sigma}_{ab} +\frac{1}{3} \widehat{\sigma}_{cd} \widehat{\sigma}{}^{cd} \widehat{h}_{ab}, \\
\widehat{H}_{ab} & = - \frac{1}{2} \widehat{\eta}_{a}{}^{cd} \widehat{D}_c \widehat{\sigma}_{d b} - \frac{1}{2} \widehat{\eta}_{b}{}^{cd} \widehat{D}_c \widehat{\sigma}_{d a},
\end{split}
\end{equation}
so if the shear tensor $\widehat{\sigma}_{ab}$ vanishes, then $\widehat{C}{}^a{}_{bcd} = 0$, i.e. $(\widehat{M}, \widehat{g}_{ab})$ is conformally flat.
\end{remark}
\section*{Tensorial obstructions for the metric to be conformal to the anisotropic fluid}
We will move now to the study of tensorial obstructions for the metric to be conformal to the one described in the previous section. \\
 Let 
\begin{equation} \label{confmet}
\widehat{g}_{ab} = e^{2 \psi} \widecheck{g}_{ab},
\end{equation}
where $\widehat{g}_{ab}$ is a metric from Theorem \ref{th1}. This section focuses only on the conformal properties of $\widecheck{g}_{ab}$, so the Einstein field equations for this metric will be imposed later on. Let
\begin{equation} \label{fourvelc}
\widehat{u}^a = e^{-\psi} \widecheck{u}^a.
\end{equation} 
The quantities describing a flow generated by $\widecheck{u}^a$ with respect to $\widecheck{\nabla}_a$ can be expressed by their counterparts related to $\widehat{u}^a$ with respect to $\widehat{\nabla}_{a}$. The Levi-Civita connections of the metrics $\widecheck{g}_{ab}$ and $\widehat{g}_{ab}$ are related to each other in a following way,
\begin{equation}
\widehat{\Gamma}^{a}{}_{bc} = \widecheck{\Gamma}{}^{a}{}_{bc} + \delta_{b}{}^a \psi_c + \delta_{c}{}^a \psi_b - \widecheck{g}_{bc} \widecheck{g}{}^{ad} \psi_d, 
\end{equation}
where $\psi_a = \partial_a \psi$, so if $\widehat{u}^a$ has a vanishing acceleration and vorticity, then
\begin{equation} \label{flowq}
\widecheck{\omega}_{ab}=0, \quad \widecheck{\sigma}_{ab} = e^{- \psi } \widehat{\sigma}_{ab}, \quad \widecheck{\theta} = e^{\psi} \widehat{\theta} - 3 \widecheck{u}{}^a \psi_a, \quad \dot{\widecheck{u}}_a = - \widecheck{D}_{a} \psi,
\end{equation}
i.e. the vorticity of $\widecheck{u}^a$ also vanishes, but the acceleration is now given by the spatial gradient of $\psi$. \\
The Cotton tensor transforms in a following way under the conformal rescaling (\ref{confmet}),
\begin{equation}
\widehat{A}_{abc} = \widecheck{A}_{abc} + \widecheck{C}_{cbad} \psi^d,
\end{equation}
whereas the Bach tensor is a conformal invariant. Together with Theorem \ref{th1}, this yields the following result.
\begin{theorem}
Let $\widecheck{g}_{ab}$ be conformal to the anisotropic fluid metric with the matter flow generated by a vector $\widehat{u}^a$ with the vanishing vorticity and acceleration and with the matter density and isotropic pressure being constant functions on the hypersurfaces orthogonal to $\widehat{u}^a$. Then 
\begin{align}
\widecheck{u}{}^a \left(\widecheck{A}_{abc}+ \widecheck{C}_{cbad} \psi^d \right)=0, \label{confob1}\\
\widecheck{h}_c{}^b \widecheck{u}{}^a \widecheck{B}_{ab} = 0.  \label{confob2}
\end{align}
\end{theorem}
This theorem can be viewed as a necessary condition for the metric to be in the conformal class of certain anisotropic fluid spacetime. It is worth to notice that although the full gradient of $\psi$ appears in (\ref{confob1}), its timelike part is not present in this equation due to the antisymmetry of the Weyl tensor in the last two indices. Thus, it is beneficial to re-express this equation in terms of the electric and magnetic parts of $\widecheck{C}_{abcd}$. We have
\begin{equation} \label{ccot}
\widecheck{u}^a \widecheck{A}_{abc} + \left( 2 \widecheck{u}_{[b} \widecheck{E}_{c]a}  + \widecheck{\eta}_{cbk} \widecheck{H}_{a}{}^k \right)  \widecheck{D}{}^a \psi = 0,
\end{equation}
where (\ref{ebweyl2}) was used. The only two independent projections of this equation can be obtained by contracting it with $\widecheck{u}{}^b \widecheck{h}_i{}^c$ or $\widecheck{h}_i{}^b \widecheck{h}_{j}{}^c$. We have
\begin{proposition}
Let 
\begin{equation}
\widecheck{u}{}^a \left(\widecheck{A}_{abc}+ \widecheck{C}_{cbad} \psi^d \right)=0,
\end{equation}
where $\widecheck{u}{}^a$ is a timelike unit vector. The only two independent projections of this equation are 
\begin{align}
\widecheck{u}{}^a \widecheck{u}{}^b \widecheck{h}_{i}{}^c \widecheck{A}_{abc} - \widecheck{E}_{i a} \widecheck{D}{}^a \psi &  =0, \label{ccot1} \\
\widecheck{u}{}^a \widecheck{h}_i{}^b \widecheck{h}_{j}{}^c \widecheck{A}_{abc}    + \widecheck{\eta}_{jik} \widecheck{H}_{a}{}^k \widecheck{D}{}^a \psi & =0 \label{ccot2}. \
\end{align}
\end{proposition}
Before moving forward we will focus on the implications of (\ref{ccot1}) and (\ref{ccot2}) for the generic case. That is, we will call $(\widecheck{M}, \widecheck{g}_{ab})$ generic if the electric part of the Weyl tensor is injective as a map from $\Gamma \left(T^* \widecheck{\Sigma} \right)$ to itself, given by $V_a \to \widecheck{E}_a{}^bV_b$. In this case, we will denote the pointwise adjugate of $\widecheck{E}_{ab}$ as $\widecheck{\widetilde{E}}_{ab}$, i.e.
\begin{equation}
\widecheck{\widetilde{E}}_{a}{}^b \widecheck{E}_{b}{}^c = ||\widecheck{E}|| \delta_a{}^c,
\end{equation}
where $||\widecheck{E}||$ is a determinant of the map $\widecheck{E}: \Gamma\left(T^* \widecheck{\Sigma} \right) \to \Gamma \left(T^* \widecheck{\Sigma} \right)$. If $(\widecheck{M}, \widecheck{g}_{ab})$ is generic, then $|| \widecheck{E}|| \neq 0$.

Applying $||\widecheck{E}||^{-1} \widecheck{\widetilde{E}}_j{}^i$ to (\ref{ccot1}) for a generic $(\widecheck{M}, \widecheck{g}_{ab})$ yields
\begin{equation}
\widecheck{D}_j \psi = \frac{\widecheck{u}{}^a \widecheck{u}{}^b \widecheck{\widetilde{E}}_j{}^c \widecheck{A}_{abc}}{||\widecheck{E}||}. 
\end{equation}
which can be used to eliminate $\widecheck{D}_a \psi$ from (\ref{ccot2}). Ultimately
\begin{equation} \label{obs}
\widecheck{u}^a  \widecheck{A}_{abc} \left(\widecheck{h}_i{}^b \widecheck{h}_{j}{}^c ||\widecheck{E}||  + \widecheck{\eta}_{jik} \widecheck{H}_d{}^k  \widecheck{u}{}^b \widecheck{\widetilde{E}}{}^{dc} \right) =0.
\end{equation}
Hence, the non-vanishing of (\ref{obs}) is an obstruction for a generic $(\widecheck{M}, \widecheck{g}_{ab})$ to be conformal to the spacetime $(\widehat{M}, \widehat{g}_{ab})$ of a certain anisotropic fluid type.
\begin{theorem}
Let $(\widecheck{M}, \widecheck{g}_{ab})$ posses a timelike unit vector field $\widecheck{u}{}^a$ orthogonal to the hypersurface $ \widecheck{\Sigma}$. Let the electric part of the Weyl tensor $\widecheck{E}_{ab}$ be an injection when treated as a map $\widecheck{E}: \Gamma \left( T^* \widecheck{\Sigma} \right) \to \Gamma \left( T^* \widecheck{\Sigma} \right)$ with a determinant $||\widecheck{E}|| \neq 0$ and a pointwise adjugate $\widecheck{\widetilde{E}}_{ab}$. Then the following expression is an obstruction for $(\widecheck{M}, \widecheck{g}_{ab})$ to be conformal to the anisotropic fluid spacetime with the flow vector $\widehat{u}^a = e^{-\psi} \widecheck{u}^a$ having vanishing vorticity and acceleration and with the density and isotropic pressure being constant on hypersurfaces orthogonal to the fluid flow lines,
\begin{equation}
\widecheck{u}{}^a  \widecheck{A}_{abc} \left(\widecheck{h}_i{}^b \widecheck{h}_{j}{}^c ||\widecheck{E}||  + \widecheck{\eta}_{jik} \widecheck{H}_d{}^k  \widecheck{u}{}^b \widecheck{\widetilde{E}}{}^{dc} \right),
\end{equation}
in the sense that its vanishing is a necessary condition for $(\widecheck{M}, \widecheck{g}_{ab})$ to be in the conformal class of such anisotropic fluid spacetime.
\end{theorem}

\section*{Anisotropic fluid spacetime in the conformal class of $(\widehat{M}, \widehat{g}_{ab})$}
Suppose now that $\widecheck{g}_{ab}$ is a solution of the Einstein field equations with
\begin{equation} \label{tchecked}
\widecheck{T}_{ab} = \left( \widecheck{\rho} + \widecheck{p} \right) \widecheck{u}_a \widecheck{u}_b + \widecheck{p} \widecheck{g}_{ab} - \widecheck{\lambda} \widecheck{\sigma}_{ab},
\end{equation}
where $\widecheck{\rho}$ and $\widecheck{p}$ are constant functions on $\widecheck{\Sigma} \perp \widecheck{u}^a$. In particular, this assumption yields:
\begin{itemize}
\item The continuity equation $\widecheck{\nabla}_a \widecheck{T}{}^{ab} = 0$, which can be split into its timelike and spacelike parts,
\begin{align}
\widecheck{\nabla}_{\widecheck{u}} \widecheck{\rho} + \left(\widecheck{\rho} + \widecheck{p} \right) \widecheck{\theta} - \widecheck{\lambda} \widecheck{\sigma}_{ab} \widecheck{\sigma}{}^{ab} =0, \label{ob0} \\
\left(\widecheck{\rho} + \widecheck{p} \right)  \dot{\widecheck{u}}_a - \widecheck{\lambda} \widecheck{D}_b \widecheck{\sigma}_a{}^b =0,  \label{ob1} 
\end{align}

\item The definition of a spatial gradient of $\widecheck{\theta}$ as a consequence of (\ref{spatth}),
\begin{equation} \label{thetagr}
  \frac{2}{3} \widecheck{D}_a \widecheck{\theta} = \widecheck{D}_b \widecheck{\sigma}{}_a{}^b - \widecheck{\sigma}_a{}^b \dot{\widecheck{u}}_b.
\end{equation}

\item The following formula for the contraction of the Cotton tensor with $\widecheck{u}{}^a$,
 \begin{equation} \label{cotacc}
\widecheck{u}^a \widecheck{A}_{abc} = \left( \widecheck{\rho} + \widecheck{p} \right)  \widecheck{u}_{[b} \dot{\widecheck{u}}_{c]} + \widecheck{\lambda}  \dot{\widecheck{u}}_{a} \psi \widecheck{\sigma}^a{}_{[b}\widecheck{u}_{c]},
\end{equation}
\end{itemize}

Since the acceleration of $\widecheck{u}^a$ is related to the spatial gradient of $\psi$,
\begin{equation} \label{accpsi}
\dot{\widecheck{u}}_a = -\widecheck{D}_a \psi,
\end{equation}
as in (\ref{flowq}), the equation (\ref{ob1}) yields
\begin{equation}  \label{ob11}
\left(\widecheck{\rho} + \widecheck{p} \right) \widecheck{D}_a \psi + \widecheck{\lambda} \widecheck{D}_b \widecheck{\sigma}_a{}^b =0.
\end{equation}
The spatial divergences $ \widecheck{D}_b \widecheck{\sigma}_{a}{}^{b}$ and $\widehat{D}_b \widehat{\sigma}_{a}{}^{b}$ are related to each other in a following way,
\begin{equation} \label{divsigs}
\widecheck{D}{}_b \widecheck{\sigma}_{a}{}^{b} = e^{ \psi } \widehat{D }_b \widehat{\sigma}_{a}{}^{b} - 3 \widecheck{\sigma}_a{}^b \widecheck{D}_b \psi,
\end{equation}
 and the latter one vanishes as in (\ref{divsig}). Hence, the equation (\ref{ob11}) reads
\begin{equation} \label{ob12}
\left(\widecheck{\rho} + \widecheck{p} \right) \widecheck{D}_a \psi - 3 \widecheck{\lambda} \widecheck{\sigma}_a{}^b \widecheck{D}_b \psi =0.
\end{equation}
Given (\ref{accpsi}) and (\ref{divsigs}), the definition (\ref{thetagr}) of the spatial gradient of $\widecheck{\theta}$ can be written as
\begin{equation} \label{ob1p}
\widecheck{\sigma}_a{}^b \widecheck{D}_b \psi + \frac{1}{3} \widecheck{D}_a \widecheck{\theta} =0,
\end{equation}
We can combine (\ref{ob12}) with (\ref{ob1p}) to get
\begin{equation} \label{pth0}
\widecheck{D}_a \left( \left(\widecheck{\rho} + \widecheck{p} \right)  \psi + \widecheck{\lambda}  \widecheck{\theta} \right)=0,
\end{equation}
The spatial derivative of the expansion $\widecheck{\theta}$ can be expressed in terms of $ \widecheck{\sigma}_{bc} \widecheck{\sigma}{}^{bc}$ using the timelike component of the matter continuity equation (\ref{ob0}). Ultimately,
\begin{equation} \label{psis2}
\widecheck{D}_a \left( \left( \widecheck{\rho} + \widecheck{p} \right)^2 \psi + \widecheck{\lambda}{}^2 \widecheck{\sigma}_{bc} \widecheck{\sigma}{}^{bc} \right)=0.
\end{equation} 
However, since $ \widecheck{\sigma}_{bc} \widecheck{\sigma}{}^{bc} = e^{2 \psi}  \widehat{\sigma}_{bc} \widehat{\sigma}{}^{bc} $, this is equivalent to
\begin{equation}
\left( \left( \widecheck{\rho} + \widecheck{p}  \right)^2 +  2 e^{2 \psi} \widecheck{\lambda}{}^2 \widehat{\sigma}_{ab} \widehat{\sigma}^{ab}  \right) \widecheck{D}_a \psi =0,
\end{equation}
where  $ \widecheck{D}_a \widehat{\sigma}_{bc} \widehat{\sigma}^{bc} = \widehat{D}_a \widehat{\sigma}_{bc} \widehat{\sigma}^{bc} = 0$ has been used (see the Proposition \ref{prop1}). This equation can be satisfied if and only if
\begin{equation} \label{constpsi}
\widecheck{D}_a \psi = - \dot{\widecheck{u}}_a = 0,
\end{equation}
i.e. the conformal factor is a constant function on $\widecheck{\Sigma}$. Hence, the acceleration $\dot{\widecheck{u}}{}^a$ vanishes.

Because of (\ref{constpsi}) and (\ref{cotacc}), the conformal Cotton (\ref{confob1}) and Bach (\ref{confob2}) equations pose no additional constraints. After combining the propagation equation for the expansion (\ref{propth}) for $\widehat{\theta}$ with its counterpart for $\widecheck{\theta}$ we arrive at
\begin{equation} \label{evolp}
3\widehat{u}{}^a \widehat{u}{}^b \widehat{\nabla}_{a}  \psi_b + \widehat{\theta} \widehat{u}^a \psi_a + e^{-2 \psi} \left(\widecheck{\Lambda} - \frac{1}{2} \left(\widecheck{\rho} + 3 \widecheck{p} \right) \right) = \widehat{\Lambda}- \frac{1}{2} \left( \widehat{\rho} + 3 \widehat{p}\right),
\end{equation}
which is a direct consequence of the double timelike projection of $\widehat{R}_{ab}$, expressed in terms of $\widecheck{R}_{ab}$.

The results of this section can be summarized by the following theorem.
\begin{theorem}
Let $(\widecheck{M}, \widecheck{g}_{ab})$ be an anisotropic fluid spacetime with the fluid density and isotropic pressure varying only along the fluid flow lines, conformal to the spacetime $(\widehat{M}, \widehat{g}_{ab})$ with the matter content of the same type and with the vanishing acceleration of the fluid, i.e. $\widehat{g}_{ab} = e^{2 \psi} \widecheck{g}_{ab}$. Assume that the fluid four-velocities are related in a following way,
\begin{equation}
\widecheck{u}^a = e^{\psi} \widehat{u}^a.
\end{equation}
Then the conformal factor $e^{2 \psi}$ is a constant function on the hypersurfaces $\widecheck{\Sigma}$ orthogonal to $\widecheck{u}{}^a$, i.e. the acceleration of $\widecheck{u}{}^a$ vanishes. Moreover, the evolution equation for $\psi$ is a direct consequence of the relation between $\widehat{R}_{\widehat{u} \widehat{u}}$ and $\widecheck{R}_{\widehat{u} \widehat{u}}$ and reads
\begin{equation}
3\widehat{u}{}^a \widehat{u}{}^b \widehat{\nabla}_{a}  \psi_b + \widehat{\theta} \widehat{u}^a \psi_a + e^{-2 \psi} \left(\widecheck{\Lambda} - \frac{1}{2} \left(\widecheck{\rho} + 3 \widecheck{p} \right) \right) = \widehat{\Lambda}- \frac{1}{2} \left( \widehat{\rho} + 3 \widehat{p}\right).
\end{equation}
 \end{theorem}
\begin{remark}
As an example of the application of the evolution equation (\ref{evolp}), we will derive the bound on the behaviour of the matter described by the 'checked' quantities if both of the metrics are conformally flat and which scale factors are reciprocal to each other, as in \cite{tod}. 

Let $\widehat{g}_{ab}$ be a vacuum solution of the Einstein field equations with positive cosmological constant $\widehat{\Lambda}$,
\begin{equation}
\widehat{g} = \frac{3}{  \widehat{\Lambda} t^2 } \left( - \mathrm{d} t^2 +\mathrm{d} x^2+ \mathrm{d} y^2 +  \mathrm{d} z^2   \right),
\end{equation}
which approximates the late stage evolution of the Robertson-Walker spacetime with vanishing curvature of the $t=$const hypersurfaces. Assume that $\widecheck{g}_{ab}$ has a similar form, but with the scale factor being an inverse of the one from $\widehat{g}_{ab}$, i.e.
\begin{equation}
\widecheck{g} = \left( \alpha^2 \frac{3}{  \widehat{\Lambda} t^2 } \right)^{-1} \left( - \mathrm{d} t^2 + \mathrm{d} x^2 + \mathrm{d} y^2 + \mathrm{d} z^2   \right), \quad \alpha=\mathrm{const},
\end{equation}
Then, from (\ref{evolp}),
\begin{equation}
 \frac{1}{2} \left(\widecheck{\rho} + 3 \widecheck{p} \right) - \widecheck{\Lambda}  =  \frac{9 \alpha^2 }{\widehat{\Lambda} t^4}.
\end{equation}
After setting $\widecheck{\Lambda} = 0 $,  $\widecheck{\rho} = 3 \widecheck{p}$ (incoherent radiation) and $\alpha^2 = \widehat{\Lambda} / \widecheck{m}$, we obtain a particular example of the duality between the two Robertson-Walker (cosmological constant - radiation) conformal cyclic cosmology aeons discussed in \cite{nurowski_meissner, nurowski2} and \cite{tod}, with the radiation from the present aeon described by the constant $\widecheck{m}$.
\end{remark}
\section*{Summary}
We investigated the freedom of choice of the conformal factor in the scenario when two spacetimes solving the Einstein field equations are conformal to each other. Both of them were assumed to have the energy-momentum tensor of the anisotropic fluid type with the constant density and isotropic pressure on the hypersurfaces orthogonal to the fluid flow lines. The four-velocities of fluids from both spacetimes were related to each other in a natural way, in accordance with the relation between their metric tensors.

As a result, we showed that the conformal factor has to be a constant function of the hypersurfaces orthogonal to the fluid flow lines. This is equivalent to the statement, that the accelerations of both four-velocities vanish. In that case, a projection of the relation between the Ricci tensors of both spacetimes along the timelike direction can be thought of as an evolution equation for the conformal factor.

\subsection*{Acknowledgements}
The author would like to thank Pawe\l \ Nurowski for suggesting this research problem, reading the manuscript and (many) helpful remarks. He would also like to thank Juan Valiene Kroon for the useful discussion. The research leading to these results has received funding from the Norwegian Financial Mechanism 2014-2021, project registration number  UMO-2019/34/H/ST1/00636.

\end{document}